December 2, 2025

# Equalizer or amplifier? How AI may reshape human cognitive differences

Maria Bigoni[1], Andrea Ichino[1,2], Aldo Rustichini[3], Giulio Zanella[1]

**Abstract**
*Machines have at times equalized physical strength by substituting for human effort, and at other times amplified these differences. Artificial intelligence (AI) may likewise narrow or widen disparities in cognitive ability. Recent evidence from the Information and Communication Technology (ICT) revolution suggests that computers increased inequality by education but reduced it by cognitive ability. Early research on generative AI shows larger productivity gains for less-skilled than for high-skilled workers. Whether AI ultimately acts as an equalizer or an amplifier of human cognitive differences is especially crucial for education systems, which must decide whether—and how—to allow students to use AI in coursework and exams. This decision is urgent because employers value workers who can leverage AI effectively rather than operate independently of it.*

Human beings differ in physical strength; yet when a weak and a strong person both use a machine such as a lever or a crane, the difference in the weight they can lift becomes much smaller than if they relied on their bare hands. Humans also differ in cognitive ability, and it is natural to ask whether Artificial Intelligence (AI) will narrow the consequences of these differences in a similar way—just as some mechanical machines have done for differences in physical strength. A large body of evidence suggests that the consequences of disparities in cognitive ability can be mitigated through effective parenting and early-life interventions in the domains of health and education (1). If such disparities remain, however, they will translate into unequal outcomes in income, consumption, and well-being. If AI could further narrow the impact of cognitive disparities, it might become one of the most powerful equalizers in human history.

There is no guarantee that this will be the case. The analogy with machines is again instructive: a hammer or a sword amplifies, rather than reduces, differences in strength—the stronger person can wield them faster, longer, and more effectively. Likewise, AI may magnify cognitive differences by enhancing the reasoning power of those who are already more capable, thereby widening the gap between individuals.

Which of these two scenarios lies ahead of us? Will AI complement or substitute human cognition? Will it exacerbate existing inequalities in initial endowments, or instead serve as an equalizer? And do the answers depend on the nature of the tasks in which humans and AI collaborate? These are among the most consequential questions for our future. To begin addressing them, we first turn to lessons from the past.

**Lessons from the ICT revolution**

The Information and Communication Technology (ICT) revolution of the 1990s offers a useful precedent for studying how technology shapes inequality. Until recently, most research on this question focused on education as the only measure of individual skills, partly because data on cognitive ability were lacking. The consensus was that the spread of ICT widened the earnings gap between college and high school graduates—a phenomenon that came to be known as "skill-biased technical change" (2).

---

[1] Department of Economics, University of Bologna, Italy.
[2] Centre for Economic Policy Research, London, UK.
[3] Department of Economics, University of Minnesota, US.

A recent study (3) extends the analysis to the differential effects of the ICT revolution among individuals with varying cognitive abilities—a novel perspective. This study uses data from *Understanding Society* to construct a representative sample of white individuals born in the UK, for whom the survey provides cognitive ability alongside education, occupation, and demographics. The inclusion of a direct measure of cognitive ability enables these researchers to investigate a key question: did ICT affect inequality across ability levels as it did across educational levels?

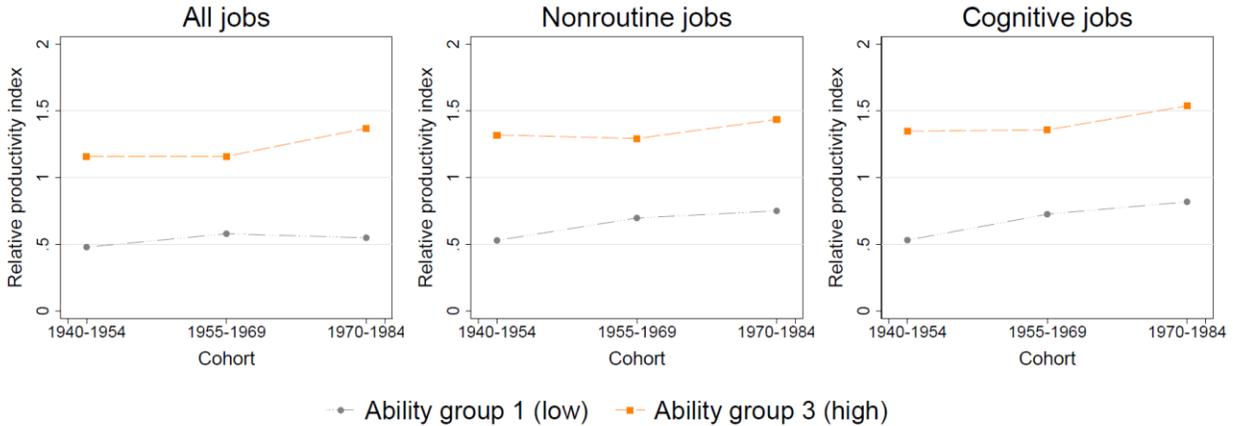

**Fig. 1. Relative productivity index of college and high school graduates by job type and by cognitive ability.** The productivity of college graduates relative to high school graduates, as measured by the ratio between the estimated coefficients of a Constant Elasticity of Substitution production function for given cognitive ability, increased over cohorts, and similarly so across high-cognitive ability individuals (above the top tercile of the cognitive ability distribution) and low-cognitive ability ones (below the bottom tercile), for all jobs and specifically for non-routine and cognitive jobs as defined in (4). These dynamics reflect the well-known phenomenon of education-biased technical change. The source of this figure is (3).

Figure 1 shows how the productivity of college graduates changed relative to high school graduates within high- and low-ability groups across three cohorts: individuals born in 1940–1954, 1955–1969, and 1970–1984. The first cohort was barely exposed to ICT, whereas the latter two were increasingly affected by computers and the Internet. Within both ability groups—especially among the high-ability—college graduates' productivity rose relative to high school graduates, and the pattern is more pronounced in non-routine and cognitive jobs as defined in (4). This pattern is consistent with previous evidence that ICT increased the relative productivity of college graduates.

This is the conclusion considering education alone. Figure 2 presents the novel perspective in (3). It shows that the same exposure to ICT reduced the productivity gap between high- and low-ability workers within each educational level, especially among college graduates in non-routine and cognitive occupations. This pattern is also evident, for example, in recent studies of surgery, where the adoption of robots has been shown to improve the performance of less skilled surgeons while diminishing the performance advantage of more skilled ones (5).

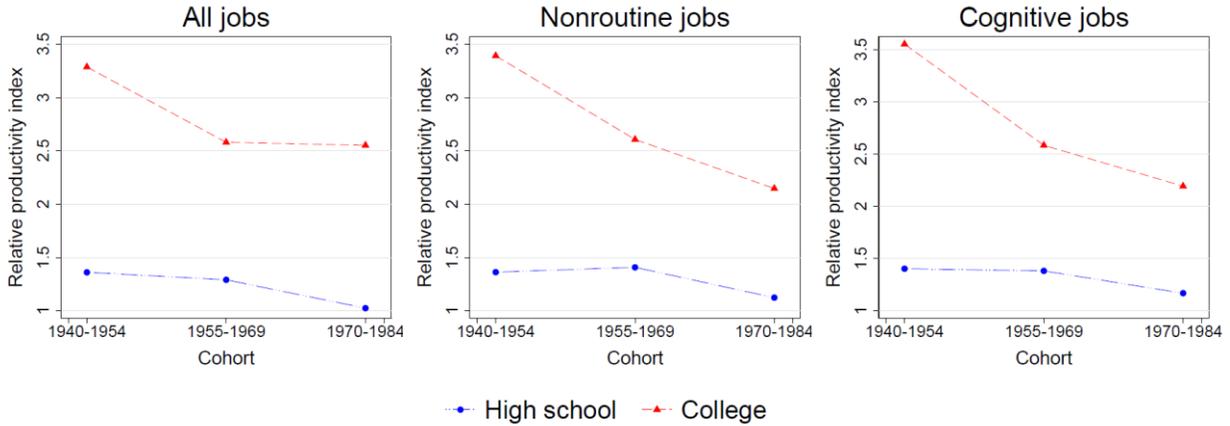

**Fig. 2. Relative productivity index of high and low cognitive ability individuals by job type and by education.** The productivity of high-cognitive ability individuals (above the top tercile of the cognitive ability distribution) relative to low-cognitive ability ones (below the bottom tercile), as measured by the ratio between the estimated coefficients of a Constant Elasticity of Substitution production function for given educational attainment, decreased over cohorts, and similarly so across college and high school graduates, for all jobs and specifically for non-routine and cognitive jobs as defined in (4). These dynamics reflect the novel phenomenon of cognitive ability-biased technical change. The source of this figure is (3).

By compressing productivity differences across ability groups within each education level, according to (3), ICT also reduced the earnings gap that typically arises from differences in cognitive ability. Figure 3 plots the present value of lifetime earnings for high school and college graduates across three ability groups. Among college graduates, earnings rose substantially more for low-ability workers and declined slightly for those in the middle and top groups. As a result, the college premium in lifetime income increased for low-ability workers but fell at the top of the ability distribution, pointing to an equalization of the income consequences of cognitive ability differences.

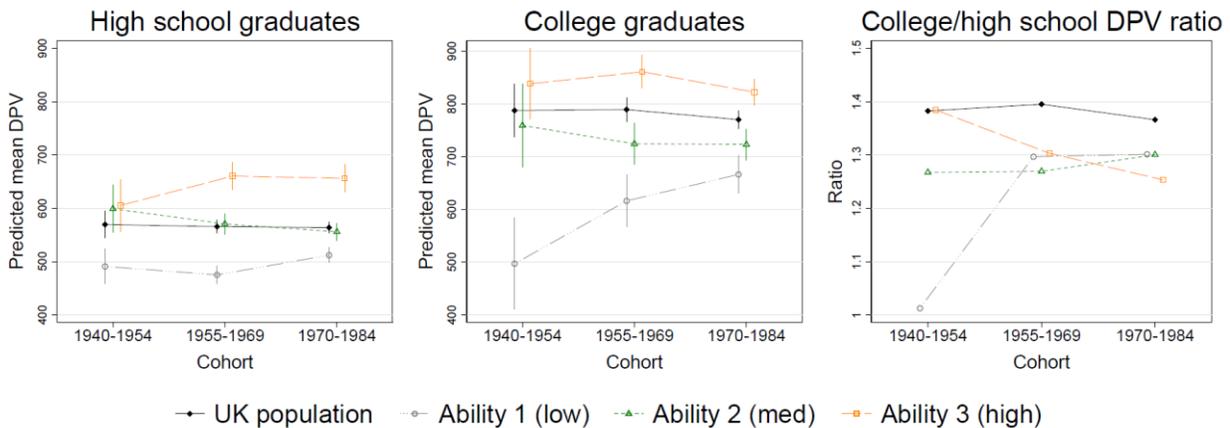

**Fig. 3. Discounted present values of lifetime earnings for high school and college graduates.** The predicted average discounted present value (DPV) of lifetime earnings, as measured by the integration of estimated earnings profiles, increased more for individuals with low cognitive ability (below the bottom tercile of the cognitive ability distribution) than for the rest, while high-cognitive-ability college graduates saw the smallest gain relative to their low-ability counterparts. Therefore, the college earnings premium increased significantly for individuals with low cognitive ability, while it decreased for their high-ability peers. The source of this figure is (3).

These results are consistent with studies showing decreasing returns to cognitive ability (6, 7) and to unobservable skills within educational levels (8), though less so with research pointing to rising wage inequality within these levels (9–11). Taken together, this evidence suggests that within each educational group, the combined effect of computers replacing workers in routine tasks and boosting productivity in nonroutine and cognitive ones may have increased the relative productivity of lower-ability individuals compared with their higher-ability peers.

**Figuring out the impact of AI**

How can the lessons from the introduction of ICT help determine whether AI will amplify or reduce the economic impact of differences in cognitive ability? The answer is not straightforward for three main reasons. First, time trends capturing exposure to ICT also reflect major cultural, social, and economic changes, making it difficult to isolate causal effects. Second, the evidence from (3) is based on a sample from a single country, whereas broader and more diverse datasets are needed. Third, even if ICT reduced inequality by ability while increasing it by education, there is no reason to expect AI to behave in the same way. The scale, speed, and nature of AI's spread through education and production make it a qualitatively different technological transformation. Still, forecasting the effects of AI is both urgent—because the transformation is already underway—and crucial—because its impact is likely to be widespread. We should therefore make the best possible use of what we have learned.

To date, only a few studies have provided direct evidence on the effects of AI on ability-related inequality. One study (12), in the context of chess learning, finds that access to AI benefits high-skill players more than low-skill ones, thereby widening the performance gap. This illustrates how AI can act as a cognitive amplifier, reinforcing pre-existing ability differences. By contrast, most other studies report the opposite. For instance, (13) shows that providing customer-support agents with a generative AI assistant substantially raises the productivity of low-experience, low-skill workers, with minimal—or even negative—effects on others. Similarly, (14) finds that access to an AI code-completion tool boosts the productivity of less able programmers. In mid-level professional writing, (15) reports that "participants with weaker skills benefited the most from ChatGPT," while in creative writing, (16) concludes that generative AI enhances individual creativity but reduces collective diversity. From a different angle, (17) shows experimentally that complementarities between workers and AI are stronger "when workers have an accurate appraisal of their own abilities," and that "people who know they have low ability gain the most from working with AI." A recent review (18) summarizes this evidence, concluding that "AI tends to most benefit entry-level workers who can quickly mimic the strategies of more experienced peers," a conclusion echoed in the theoretical analysis of (19).

Overall, these early findings suggest that AI may act as a substitute for human cognition, narrowing performance gaps from the top rather than amplifying them. Yet some additional evidence points to a different form of equalization—one that levels downward. In a recent experiment, (20) shows that when people use ChatGPT for essay writing, brain activity in areas linked to attention and working memory declines. The authors describe this as a "cognitive debt," implying that large language models may equalize performance by reducing everyone's level of engagement. Similarly, (21) finds that unrestricted access to ChatGPT improves performance on practice exercises but harms later exam results once AI is withdrawn, revealing a risk of downward leveling. Laboratory evidence (22) echoes this pattern: access to ChatGPT helps weaker students catch up in a programming course, but mainly by reducing cognitive effort—suggesting a short-term equalization that may come at the cost of deeper learning. These results resonate with Jonathan Haidt's argument (23) that the digital transformation of childhood has constrained the development of attention and deep reasoning, replacing unstructured play, face-to-face interaction, and boredom-driven imagination with continuous digital stimulation through smartphones, social media, and possibly AI. Others (24), however, caution against drawing direct causal links, noting that most evidence on these effects remains small and inconsistent.

In summary, the evidence on whether AI equalizes the impact of cognitive ability—and, if so, in which direction—is still fragmentary and inconclusive.

**Answering this question is particularly important for educational institutions**

Such inconclusiveness makes an answer to the question of how AI will ultimately affect cognitive differences all the more urgent, especially in education systems worldwide. These systems must *now* decide whether—and how—to integrate AI into teaching and assessment. Should students be allowed to use AI in coursework, and even during exams?

In the labor market, firms increasingly expect workers to use AI. Employers care about the productivity of workers who can effectively leverage AI, not of those who are prevented from doing so. Ideally, they want employees who know when to rely on AI—and when to question its advice. This implies that education systems should learn to evaluate and rank students *with* AI, not *without* it. A pure ability ranking matters less if AI systematically alters performance and compresses the impact of individual ability differences. And if AI not only modifies the effects of ability differences but also reshuffles individual ranks, then it becomes even more important that schools assess students in the presence of AI, rather than in its absence.

A subtler interaction between AI and human cognition concerns creativity. Cognitive ability among knowledge workers exhibits a discontinuity: ordinary individuals routinely apply existing knowledge to answer questions; in contrast, geniuses generate new knowledge in creative ways. Recent theoretical work (25) suggests that the labor market may absorb the AI-induced expansion of cognitive capacity by displacing routine knowledge workers, while highly creative individuals focus on questions that for the moment remain beyond the reach of AI. The limited empirical evidence (26, 27) so far indicates that AI does not yet outperform humans in creative tasks. In this scenario, AI could amplify differences in a more radical, indirect way—by reallocating tasks toward a minority of highly creative individuals. It follows that education and training should now prioritize the development of creativity with access to AI, rather than the mere application of existing knowledge in isolation from it, as has traditionally been done.

**The research effort ahead**

In light of these considerations, designing and implementing comprehensive field experiments in education should be a priority. Such studies should test how human cognitive ability and AI interact across different tasks, and how this interaction affects performance in learning. This research will require close collaboration among economists, psychologists, educators, computer scientists, and policymakers.

A basic experimental design could proceed as follows. After measuring participants' cognitive ability, assign them to three groups. The control group would complete tasks without AI assistance. In the first treatment arm, participants would use AI under strict time constraints; in the second, they would have unlimited time to interact with AI. This dual-treatment setup reflects that time pressure is a key feature of real-world learning and work, where AI may either complement or substitute human cognition.

Our prior is that time-constrained access to AI will shift the performance distribution to the right, but more so for high-ability individuals. AI helps most when users can pose the right questions and verify the answers—skills that require both time and cognitive resources. Under tight time constraints, lower-ability participants may struggle to use AI effectively or even perform worse than without it. When time limits are relaxed, by contrast, AI may enable lower-ability participants to close the gap, compressing performance differences from below.

A more challenging experimental design, aimed at testing specifically how AI interacts with human creativity, should assign participants to inventive tasks that require the creation of new knowledge, such as research that advances scientific or technological understanding. A significant challenge in this experimental setup lies in developing a rigorous and credible system for evaluating the quality of participants' responses. Assessing creative output requires evaluators with high professional expertise, which makes it difficult to recruit and motivate suitable raters. A promising solution might therefore combine human and AI evaluation: (i) the AI system could provide an initial estimation of the quality across the full set of participants' responses, and (ii) a selected subset of both responses and AI rankings could then be independently evaluated by expert human judges, serving as a calibration or validation mechanism for the AI-based assessment.

Envisioning such experiments is relatively easy; implementing them is not. The tasks must be realistic and meaningful, raising both ethical and logistical challenges. Schools and universities may hesitate to expose students to experimental variation in AI access, yet only such evidence can reveal whether students should be allowed to use AI and be evaluated on their joint productivity with it.

Finally, this research effort will require substantial funding, the source of which remains uncertain. But there is no other reliable way to determine whether AI narrows or widens cognitive disparities and how education and policy should respond.

**Acknowledgements**

Portions of the text were revised for clarity and grammar using OpenAI's ChatGPT (GPT-5, 2025 version). The authors reviewed all AI-assisted edits and are solely responsible for the final content.